\documentstyle[12pt]{article}
\newcommand{\beq}{\begin{equation}}

\newcommand{\eeq}{\end{equation}} 

\newcommand{\bra}{\begin{array}}

\newcommand{\era}{\end{array}}

\newcommand{\al}{\alpha}

\newcommand{\Ga}{\Gamma}

\newcommand{\Om}{\Omega}
\newcommand{\de}{\delta}

\newcommand{\Te}{\Theta}

\newcommand{\ot}{\otimes}

\begin{document}
\begin{center}

{\Large\bf Multi-Species Anyons Supersymmetry on Two-Dimensional Space}
\vskip1.2truecm
{\large\bf Jamila Douari}
\footnote
{\null\hspace{0.5cm} E-mail: douari@sun.ac.za}
\vskip.5truecm
Universit\'e Mohammed V, Facult\'e des Sciences, D\'epartement de Physique, LPT-ICAC, Av. Ibn Battouta, B.P.1014, Agdal, Rabat, MOROCCO.

\end{center}

\hoffset=-1cm\textwidth=20cm\pagestyle{empty}
\vskip1.4truecm

\begin{abstract}
The algebra of multi-species anyons characterized by different statistical parameters $\nu_{ij}=e_{i}e_{j}\Phi_{i}\Phi_{j}/(2\pi)$, $i,j=1,...,n$ is redefined by basing on fermions and $k_{i}$-fermions ($k_{i}\in\bf{N}\rm /\{0,1\}$ with $i\in\bf{N}\rm $) and its superalgebra is constructed. The so-called fractional supersymmetry of multi-species anyons is realized on 2d lattice.
\end{abstract}

\section{Introduction}
\hspace{.3in}The idea of supersymmetry (SUSY) has stimulated
new approaches in many branches of physics. An evidence example has
been found for a dynamical SUSY related even-even and even-odd nuclei
\cite{1,2}. SUSY is a theoretically attractive possibility for several
reasons. It is the unique possibility for non-trivial extension of the known
symmetries of space and time \cite{3}. Many Physicists have developed theories of
SUSY, particularly in the context of Grand Unified Theories, which successfully
attempt to combine the strong and electroweak interactions \cite{4}.

In SUSY quantum mechanics one is considering a simple realization of SUSY
algebra, involving the fermionic and bosonic operators \cite{5}, which had to move
beyond Lie algebra to "graded" Lie algebras. Graded Lie algebras are just
like Lie algebras except they use anti-commutation and commutation relations.

In view of the fact that the SUSY provided us with an elegant symmetry
between fermions and bosons, it was natural to enquire if there exists a
generalization which includes the exotic statistics. Various kinds of such
extensions were realized: paraSUSY (with parafermions), fractional SUSY
(using the $q$-bosons with $q$ a root of unity) and nonlinear SUSY
(bosonization of supersymmetric quantum mechanics) [6-14]. Another
generalization can be treated concerning the case of anyons. An attempt in
this sense is considered, in this work, combining two different kinds of
anyons.

On other had, during roughly 20 years, anyons have attracted a great attention to understand the physics of lower dimensions. Wilczek is generally
credited to others earlier \cite{15,16}. He defined these new particles as being a vortex of the gauge-field which are the intersection of a plane and tubes of magnetic flux electrically charged \cite{17}. Owing the work \cite{18} of
Leinaas and Merhyeim on identical particles, we all know that anyons have
a lot to do with braids. Thus, the quantum algebras seems to be a good condidates describing their symmetries \cite{19,20,21}. The realization of these algebras had been
started by Lerda and Sciuto \cite{22} and others.

Now, let us pay close attention to extend the notion of symmetry for these
new systems. So, what about ``super''-symmetry (SUSY) of nyons? and what
kind of SUSY could exist in 2d space? In our present work, we will consider
a system of n types of anyons (those we call multi-species anyons \cite{21}) characterized by different statistical parameters denoted
$$\nu_{ij}=e_{i}e_{j}\Phi_{i}\Phi_{j}/(2\pi),\phantom{~~~~~}i,j=1,...,n$$
$e_i$ is charge of the $i^{th}$ particle and $\Phi_i $ is its flux. Each statistical parameter $\nu_{ij}$ describes the interaction of the $i^{th}$ anyon and the $j^{th}$ one and defines the existence of different fractional statistics.

In this context, this work will discuss the construction of SUSY describing a
system of multi-species anyons. First, we redefine the algebra of anyons by basing on fermions and $k_i$-fermions \cite{23,24} ($k_i \in\bf N\rm/{0,1}$) i 2d square lattice. Let us denote here that the spatial coordinates $x$ are restricted to a multiple of a lattice spacing $a$,  i.e. $x=na$ with $n$ an integer. By removing the lattice structure the algebraic results don't change. Second, we generalize the deinition of anyonic algebra by taking into account all kind of anyons. So we construct the "super"-algebra associated to our system. By introducing supercharges in terms of two different anyonic operators we realize the SUSY of multi-species anyons system which can be called as fractional one since the construction is based on the nature of anyons.

This paper is organized as follows: In section 2, we introduce the definition of multi-species anyonic oscillators and their algebras on 2d square lattice based on fermionic ones. In section 3, we define the $k_i$-fermions and we extend the Lerda-Sciuto definition to construct the multi-species $k_i$-fermionic anyons and their algebras. In section 4, we construct the anyonic superalgebra by considering a system of different species of anyons. In section 5, we use the generators of anyonic superalgebra to construct the supercharges of multi-species anyons SUSY. In section 6, we discuss the irreducible representations of anyonic algebras and superalgebra. In the last section, we summarize the main result of the paper.

\section{Multi-species anyonic oscillators}
\hspace{.3in}Let $\Om$ be a $2d$ square lattice with spacing $a=1$. We give a
two-component fermionic spinor field by
\beq
S^- =\pmatrix{s_1 ^- (x)\cr s_2 ^- (x)\cr},
\eeq
and its conjugate hermitian by
\beq
S^+=(s_1 ^+ (x),s_2 ^+ (x)),
\eeq
such that the components of these fields satisfy the following standard
anti-commutation relations
\beq
\bra{rcl}
\lbrace s_i ^- (x),s_j ^- (y)\rbrace &=&0\\
\lbrace s_i ^+ (x),s_j ^+ (y)\rbrace &=&0\\
\lbrace s_i ^- (x),s_j ^+ (y)\rbrace &=&\delta_{ij}\delta(x,y),
\era
\eeq
$\forall i,j\in\{1,2\}$ and $\forall x,y\in\Omega$. Here, $\delta(x,y)$ is the
conventional lattice $\delta$-function: $\delta(x,y)=1$ if $x=y$ and vanishes
if $x\ne y$.

The expression of anyonic oscillators are given in terms of fermionic spinors as follows
\beq
\bra{rcl}
b_{ij} ^- (x_{\pm})&=&e^{i\nu_{ij}\Delta_i (x_{\pm})}s_i ^- (x)\\
b_{ij} ^+ (x_{\pm})&=&s_i ^+ (x)e^{-i\nu_{ij}\Delta_i (x_{\pm})},
\era
\eeq
where $\nu_{ij}$ are called statistical parameters and the
elements $\Delta_i (x_{\pm})$ are given by
\beq
\Delta_i (x_{\pm})=\sum\limits_{y\in\Omega}s_i ^+
(x)\Theta_{\pm\Gamma_{x}}(x,y)s_i ^- (y),
\eeq
with $\Theta_{\pm\Gamma_{x}}(x,y)$ are the so-called angle functions and its definition o 2d square lattice was recited in the references \cite{12} and \cite{13}, where $\Gamma_{x}$ is the curve associated to each site $x\in\Om$ and the signs + and - indicates the two kinds of rotation direction on $\Om$.

The elements $\Delta_i (x_{\pm})$ satisfy the following commutation relations
$$[\Delta_i (x_\pm),\ s_j ^-
(y)]=-\delta_{ij}\Theta_{\pm \Gamma_x}(x,y) s_i ^- (y)$$ $$[\Delta_i (x_\pm),\
s^+ _j (y)]=\delta_{ij}\Theta_{\pm \Gamma_x}(x,y)s^+ _- (y)$$ $$[\Delta_i
(x_\pm),\ \Delta_j (y_\pm)]=0$$
Now, we can show that the anyonic oscillators satisfy the following algebraic relations
\beq
\bra{llllllllll}
\lbrack b_{ij} ^- (x_{\pm}),b_{ik} ^- (y_{\pm})\rbrack_{\Lambda_{ijk}^\mp}=0,& x>y\\
\lbrack b_{ij} ^- (x_{\pm}),b_{ik} ^+ (y_{\pm})\rbrack_{\Lambda_{ijk}^\pm}=0,& x>y\\
\lbrack b_{ij} ^+ (x_{\pm}),b_{ik} ^- (y_{\pm})\rbrack_{\Lambda_{ijk}^\pm}=0,& x>y\\
\lbrack b_{ij} ^+ (x_{\pm}),b_{ik} ^+ (y_{\pm})\rbrack_{\Lambda_{ijk}^\mp}=0,& x>y\\
\lbrack b^{-}_{ij}(x_{\pm}),b^{+}_{ik}(x_{\pm})\rbrack =1,& \\
\lbrack b_{ij} ^- (x_{\pm}),b_{kl} ^+ (y_{\pm})\rbrack=0,& i\ne j\\
\lbrack b_{ij} ^+ (x_{\pm}),b_{kl} ^- (y_{\pm})\rbrack=0,& i\ne j\\
\lbrack b_{ij} ^+ (x_{\pm}),b_{kl} ^+ (y_{\pm})\rbrack=0,& i\ne j\\
\lbrack b_{ij} ^{\pm} (x_{-}),b_{kl} ^{\pm} (y_{+})\rbrack=0,& \forall i,j,k,l\\
\lbrack b_{ij} ^- (x_{-}),b_{kl} ^+
(y_{+})\rbrack=\delta_{ik}\delta_{ij}\delta(x,y)\Gamma_{ijk}^{s_i ^+ (z)s_i ^- (z)},&\\
\lbrack b_{ij} ^- (x_{-}),b_{kl} ^+ (y_{+})\rbrack= \delta_{ik}\delta_{ij}\delta(x,y)\Gamma_{ijk}^{-s_i ^+ (z)s_i ^- (z)}.&
\era
\eeq
where
$$
\begin{array}{lll}
\Lambda_{ijk}^\pm =e^{\pm
i(\nu_{ij}\Theta_{-\Ga_{x}}(x,y)-\nu_{ik}\Theta_{+\Ga_{y}}(y,x))},\\
\Gamma_{ijk}=e^{i\sum\limits_{z\ne
x}(\nu_{ij}\Theta_{-\Ga_{x}}(x,z)-\nu_{ik}\Theta_{+\Ga_{y}}(z,x))},\\
\lbrack X,Y \rbrack_{\Lambda} = XY + \Lambda YX
\end{array}
$$
and $$x>y\Leftrightarrow
\left\{\begin{array}{l}
x_+>y_+\Leftrightarrow\left\{\begin{array}{ll}
x_2>y_2\\[2mm]
x_1>y_1, x_2 =x_1\end{array}\right.\\[3mm]
x_-<y_-\Leftrightarrow
\left\{\begin{array}{l}
x_2<y_2\\[2mm]
x_1<y_1, x_1 =x_2\end{array}\right.\end{array}\right. $$
One obtains also
\beq
(b_{ij} ^\pm (x_\pm))^2 =0,
\eeq
which is known as the hard core condition.

Let us remark that if we suppose $i=j$ in the expression of $\nu_{ij}$ given in section 1, we refind the algebraic relations of anyons constructed by Lerda and Sciuto in \cite{12}. Also, we would like to stress that despite the deformation of our above algebraic relations, the anyonic oscillators don't have anything to do with the $k_i$-fermions which have deformed algebraic relations (will be discussed in the next section) for several reasons: (i) the $k_i$-fermions can be defined in any dimensions whereas the ayons are srictly two-dimensional objects, (ii) the ayons are non-local contrary to the $k_i$-fermions constitute a mathematical, introduces in the context of quantum algebras, which is used to go beyond the conventional statistics in any dimension and can take into account some perturbation (deformation) responsible of small deiations from the Fermi-Dirac and Bose-Einstei usual statistics.
\section{Multi-species $k_i$-fermionic anyons on 2d square lattice}
\hspace{.3in}In this part of work, we will costruct the $k_i$-fermionic anyons on 2d square lattice $\Om$ from the $k_i$-fermions.

To define the $k_i$-fermionic anyons, we extend the Lerda and Sciuto definition
as
\beq
\bra{rcl}
a^- _{ij} (x_{\pm})&=&e^{i\nu_{ij}D_i (x_\pm)}f^- _i (x)\\
a^+ _{ij} (x_{\pm})&=&f^+ _i (x)e^{-i\nu_{ij}D_i (x_\pm)},
\era
\eeq
where $D_i (x_\pm)$ is given by
\beq
\sum\limits_{y\in\Omega}\Theta_{\pm\Gamma_{x}}(x,y)N_i (y),
\eeq
$N_i (y)$ is the number operator of $k_i$-fermmions on 2d square lattice defined by $f^- _i (x)$ and $f^+ _i (x)$ the $k_i$-fermmionic annihilation and creation operators respctivel as follows
$$
\bra{rcl}
f^+ _i (x)f^- _i (x)=[N_i (x)]_{q_i },\\
f^- _i (x)f^+ _i (x)=\bf 1\rm +[N_i (x)]_{q_i },
\era
$$
with $\bf 1\rm $ the identity, $q_i =^{i\frac{2\pi}{k_i}}$ $(k_i \in\bf N\rm/{0,1}, i\in\bf N\rm)$ and the noton $[x]_q =\frac{q^x -1}{q-1}$.

The $k_i$-fermmionic operators satisfy the algebraic relations
\beq
\bra{lllllll}
\lbrack f_{i}^- (x),f_{j}^{+}(y)\rbrack_{q_{i}^{\delta_{ij}}} & = &
\delta_{ij}\delta(x,y) \\
\lbrack f_{i}^- (x),f_{j}^- (y)\rbrack_{q_{i}^{\delta_{ij}}} & = &
\begin{array}{cc}
0 & \mbox{$\forall x,y$, $\forall i,j$}
\end{array}
\\
\lbrack f_{i}^{+}(x),f_{j}^{+}(y)\rbrack_{q_{i}^{\delta_{ij}}} & = &
\begin{array}{cc}
0 & \mbox{$\forall x,y$, $\forall i,j$}
\end{array}
\\
\lbrack N_{i}(x),f_{j}^- (y)\rbrack & = & -\delta_{ij}\delta(x,y)f_{i}^- (x) \\
\lbrack N_{i}(x),f_{j}^{+}(y)\rbrack & = & \delta_{ij}\delta(x,y)
f_{i}^{+}(x) \\
(f_{i}^- (x))^{k_{i}}=(f_{i}^{+}(x))^{k_{i}}=0.
\era
\eeq

By using the previous tools, the operators $a_i ^\pm (x_\pm)$ constructed
from $k_i$-fermionic oscillators satisfy the following algebraic relations
\beq
\bra{lllllllll}
\lbrack a_{ij} ^- (x_{\pm}),a_{ik} ^- (y_{\pm})\rbrack_{p^\mp _{ijk}}=0,& x>y\\

\lbrack a_{ij} ^- (x_{\pm}),a_{ik} ^+ (y_{\pm})\rbrack_{p^\pm _{ijk}}=0,& x>y\\

\lbrack a_{ij} ^+ (x_{\pm}),a_{ik} ^- (y_{\pm})\rbrack_{p^\pm _{ijk}}=0,& x>y\\

\lbrack a_{ij} ^+ (x_{\pm}),a_{ik} ^+ (y_{\pm})\rbrack_{p^\mp _{ijk}}=0,& x>y\\

\lbrack a_{ij} ^- (x_{\pm}),a_{kl} ^+ (y_{\pm})\rbrack=0,& i\ne k\\

\lbrack a_{ij} ^+ (x_{\pm}),a_{kl} ^- (y_{\pm})\rbrack=0,& i\ne k\\

\lbrack a_{ij} ^+ (x_{\pm}),a_{kl} ^+ (y_{\pm})\rbrack=0,& i\ne k\\

\lbrack a_{ij} ^{\pm} (x_{-}),a_{kl} ^{\pm} (y_{+})\rbrack=0,& \forall i,j\\

\lbrack a_{ij}^{-}(x_{\pm}) ,a_{ik}^{+}(x_{\pm}) \rbrack_{q_{i}} =1,&\\

\lbrack a^{\pm}_{ij}(x_{-}),a^{\pm}_{kl}(y_{+})\rbrack_{q_{i}} =0,&
 x,y\in \Omega,\\  
\\
\lbrack a^{-}_{ij}(x_{-}),a^{+}_{kl}(y_{+})\rbrack_{q_{i}}=\de_{ik} \de(x,y)\Gamma_{ijk}^{N_{i}(z)},& \forall
i,j,k=1,...,n.\\
\\
\lbrack a^{-}_{ij}(x_{+}),a^{+}_{kl}(y_{-})\rbrack_{q_{i}}=\de_{ik} \de(x,y)\Gamma_{ijk}^{-N_{i}(z)},& \forall
i,j,k=1,...,n.
\era
\eeq
$\forall i,j,k,l=1,...,n$. In this equation
$$p^- _{ijk}=q_{i}e^{-i(\nu_{ij}\Theta_{-\Ga_{x}}(x,y)-\nu_{ik}\Theta_{+\Ga_{y}}(y,x))}$$
and
$$p^+ _{ijk}=q_{i}e^{i(\nu_{ij}\Theta_{-\Ga_{x}}(x,y)-\nu_{ik}\Theta_{+\Ga_{y}}(y,x))},$$
and $q_{i}=e^{i\frac{2\pi}{k_i}}$ ($k_{i}\in\bf{N^*}\rm$).

We also have the following nilpotency condition
\beq
(a^{\pm}_{ij}(x_{\pm}))^{k_i} =0
\eeq
which can be interprated as a hard core condition generalizing the Pauli
exclusion principle. In the particular case $k_i =2$ (undeformed ferions), we recover the multi-species anyonic algebra of secton 2.
\section{Anyonic Superalgebra}
\hspace{.3in}In this section, we will consider $n$ species of $k_i$-fermionic anyons ($i=0,1,...,n-1$) having
different fractional spin and characterized by different fractional statistical
parameters $\nu_{ij}$. To construct the associate algebra the new generators
will be defined as direct sum of $k_i$-fermionic anyons oscillators given
by the equations (8). This definition will be in a cyclic order to take into
account all kind of anyons can exist in the combined system. So, the
constructed algebra will be in a "graded" form, and we will call it anyonic
superalgebra. We define its generators as follows
\beq
\bra{rcl}
A_{ij}^- (x_{\pm})&=a_{ij}^- (x_{\pm})\oplus a_{i+1,j}^- (x_{\pm})\oplus ...\oplus a_{i+(n-1),j}^- (x_{\pm})\\
\\
A_{ij}^{+}(x_{\pm}) &=a_{ij}^{+}(x_{\pm})\oplus a_{i+1,j}^{+}(x_{\pm})\oplus ...\oplus a_{i+(n-1),j}^{+}(x_{\pm}).
\era
\eeq

In a straightforward calculation, we prove that these operators obey to the following commutation relations
\beq
\bra{rcl}
\lbrack A_{ij}^- (x_{\pm}) ,A_{ik}^- (y_{\pm}) \rbrack_{P_{ijk}^{\mp}} =0,& x>y\\
\\
\lbrack A_{ij}^{+}(x_{\pm}) ,A_{ik}^{+}(y_{\pm}) \rbrack_{P_{ijk}^{\mp}} =0,& x>y\\
\\
\lbrack A_{ij}^- (x_{\pm}) ,A_{ik}^{+}(y_{\pm}) \rbrack_{P_{ijk}^{\pm}} =0,& x>y\\
\\
\lbrack A_{ij}^{+}(x_{\pm}) ,A_{ik}^- (y_{\pm}) \rbrack_{P_{ijk}^{\pm}} =0,& x>y\\
\\
\lbrack A_{ij}^- (x_{\pm}) ,A_{ik}^{+}(x_{\pm}) \rbrack_{Q_{i}} =1\!1\ ,& \\
\\
\lbrack A_{ij}^{\pm}(x_{-}) ,A_{kl}{\pm}(y_{+}) \rbrack_{Q_{i}} = 0,& \forall x,y\in\Om\\
\\
\lbrack A_{ij}^- (x_{-}) ,A_{kl}^{+}(y_{+}) \rbrack_{Q_{i}} =\de_{ik}
\de(x,y)\Ga_{[ijk]}, & \forall i,j,k,l=1,...,n.\\
\\
\lbrack A_{ij}^- (x_{+}) ,A_{kl}^{+}(y_{-}) \rbrack_{Q_{i}} =\de_{ik}
\de(x,y)\Ga_{[ijk]}^{-1}, & \forall i,j,k,l=1,...,n.
\era
\eeq
Let us denote here that $Q_{i}^{k}= 1\!1\ $, $k=k_{0}k_{1}...k_{n-1}$. The new
operators  $A_{ij}^- (x_{\pm})$ and $A_{ij}^{+}(x_{\pm})$ also satisfy the
following nilpotency condition
\beq
\bra{rl}
(A_{ij}^- (x_{\pm})) ^{k}=(A_{ij}^{+}(x_{\pm})) ^{k}=0, & k \in{\bf N^*}
\era
\eeq
with $k=k_{0}....k_{n-1}$, and $\forall i,j,k=1,...,n$, with
\beq
\bra{ll}
\Ga_{[ijk]}= \pmatrix{e^{i\sum\limits_{z\ne
x}(\nu_{ij}\Theta_{-\Ga_{x}(x,z)}-\nu_{ik}\Theta_{+\Ga_{x}(x,z)})N_{i}(z)}&&\cr
&\ddots&\cr
&&e^{i\sum\limits_{z\ne
x}(\nu_{i+(n-1),j}\Theta_{-\Ga_{x}(x,z)}-\nu_{i+(n-1),k}\Theta_{+\Ga_{x}(x,z)})N_{i}(z)}\cr}.\\
\\
P_{ijk}^- =\pmatrix{p_{ijk}^- &&&& \cr
&p_{i+1,jk}^- &&&\cr
&&p_{i+2,jk}^- &&\cr
&&&\ddots&\cr
&&&&p_{i+(n-1),jk}^- \cr}\\
\\
P_{ijk}^+ =\pmatrix{p_{ijk}^+ &&&&\cr
&p_{i+1,jk}^+ &&&\cr
&&p_{i+2,jk}^+ &&\cr
&&&\ddots&\cr
&&&&p_{i+(n-1),jk}^+ \cr}\\
\\
Q_{i}=\pmatrix{q_{i}&&&&\cr
&q_{i+1}&&&\cr
&&q_{i+2}&&\cr
&&&\ddots&\cr
&&&&q_{i+n-1}\cr}
\era
\eeq
The equation (15) generalizes the hard core condition for combined anyonic
system. This means that no more that $(k_i -1)$ particles can live in the
same state of anyons constructed from $k_i $-fermions $(i=0,1,...,n-1)$.
\section{Multi-species $k_i$-fermionic Anyons Supersymmetry}
\hspace{.3in}In this section, We will consder a combined system of multi-species $k_i $-fermionic anyons. We define the supercharges of the SUSY associated to the present system in terms of the generators of anyonic superalgebra given in section 4 as follows
\beq
\bra{rcl}
C^{--}_{ijrs} (x)&= A_{ij}^- (x_{-}) A_{rs}^{+}(x_{+})\\
C^{+-}_{ijrs} (x)&= A_{ij}^{+}(x_{+}) A_{rs}^- (x_{-})
\era
\eeq
with $i,j,r,s=1,...,n$ and $i\ne r$.

Using the above tools, these new supercharges obey to the following commutation relation
\beq
Q_{i}C^{+-}_{ijrs}(x)C^{--}_{ijrs}(y)-Q_{r}C^{--}_{ijrs}(y)C^{+-}_{ijrs}(x)=
\de(x,y)[B_{r}Q_{i}[\aleph_{i}(x)]_{Q_{i}}-B_{i}Q_{r}[\aleph_{r}(x)]_{Q_{r}}]
\eeq
with
\beq
\bra{ll}
B_{i}&= \pmatrix{d_{ij}^{(\sum\limits_{z>x}-\sum\limits_{z<x})N_{i}(z)} &&&\cr
& d_{i+1,j}^{(\sum\limits_{z>x}-\sum\limits_{z<x})N_{i+1}(z)} &&\cr
&& \ddots&\cr
&&& d_{i+n-1,j}^{(\sum\limits_{z>x}-\sum\limits_{z<x})N_{i+n-1}(z)}\cr}\\
\\
\lbrack\aleph _{i}(x)\rbrack_{Q_i}&= \pmatrix {[N_{i}(x)]_{q_i} &&&\cr
& [N_{i+1}(x)]_{q_{i+1}} &&\cr
&& \ddots&\cr
&&& [N_{i+n-1}(x)]_{q_{i+n-1}}\cr} \\
\\
d_{ij}&=e^{i\nu_{ij}\pi},
\era
\eeq
here $N_{i}(x)= a_{ij}^{+}(x_{\pm}) a_{ij}^- (x_{\pm})$.

To have an invariant expression under the hermitian conjugate we are doing
the following step by computing the hermitian conjugate of Eq.(18)
\beq
\bra{ll}
Q_{i}^{-1}C^{-+}_{ijrs}(y)C^{++}_{ijrs}(x)-Q_{r}^{-1}C^{++}_{ijrs}(x)C^{-+}_{ijrs}(y)=\\
\phantom{~~~~~~~~~~~~~~~~~}\de(x,y)
[B_{r}^{-1}Q_{i}^{-1}[\aleph_{i}(x)]_{Q_{i}^{-1}}-[B_{i}^{-1}Q_{r}^{-1}[\aleph_{r}(x)]_{Q_{r}^{-1}}],
\era\eeq
where $C^{\pm +}_{ijrs}(x)$ are the hermitian conjugates of the generalized
supercharges $C^{\pm -}_{ijrs}(x)$. We write
\beq
\bra{rcl}
C^{-+}_{ijrs}(x)&= A_{rs}^- (x_{+}) A_{ij}^{+}(x_{-})\\
\\
C^{++}_{ijrs}(x)&= A_{rs}^{+}(x_{-}) A_{ij}^- (x_{+}).
\era
\eeq
Let us remark here that these generators satisfy
\beq
\bra{rl}
(C^{\pm +}_{ijrs}(x)) ^{k}=(C^{\pm -} _{ijrs}(x)) ^{k}=0,& k=k_{0}k_{1}...k_{n-1}.
\era
\eeq

Now we introduce a hermetic operator denoted $H(x)$ as the sum of the
equalities (18) and (20), then we get
\beq
\bra{rcl}
H(x)&=&B_{r}Q_i [\aleph_i (x)]_{Q_i}-B_i Q_{r}[\aleph_r (x)]_{Q_{r}}\\
\\
&&+B_{r}^{-1}Q_{i}^{-1} [\aleph_i (x)]_{Q^{-1}_{i}}-B^{-1}_{i} Q_{r}^{-1}[\aleph_r (x)]_{Q_{r}^{-1}}.
\era
\eeq
In a straightforward computation we get the following relation
\beq
[N_i (x)]_{q_i ^{-1}}=q_i ^{1-N_i (x)}[N_i (x)]_{q_i},
\eeq
then the operator $[\aleph_i (x)]_{Q^{-1}_{i}}$ (Eq.(19)) can be written as
\beq
\bra{rcl}
[\aleph_i (x)]_{Q^{-1}_{i}}&=&\pmatrix {q_i ^{1-N_i (x)} &&&\cr
& q_{i+1} ^{1-N_{i+1}(x)} &&\cr
&&\ddots&\cr
&&& q_{i+n-1} ^{1-N_{i+n-1} (x)}\cr} [\aleph_i (x)]_{Q_i}\\
\\
&=&A_i [\aleph_i (x)]_{Q_i}
\era 
\eeq
Thus the equality (23) will be rewritten as
\beq
\bra{rcl}
H(x)&=&(B_{r}Q_i +B_{r}^{-1}J_{i})[\aleph_i (x)]_{Q_i}\\ \\
&&-(B_i Q_{r}+B^{-1}_{i} J_{r})[\aleph_{i+1} (x)]_{Q_{i+1}},
\era
\eeq
where
\beq
J_{i}=\pmatrix {q_i ^{-N_i (x)} &&&\cr
& q_{i+1} ^{-N_{i+1}(x)} &&\cr
&&\ddots&\cr
&&& q_{i+n-1} ^{-N_{i+n-1} (x)}\cr}.
\eeq
To extend these results to 2d continuum space, it is sufficient to summon on
all the sites of 2d lattice
\beq
\bra{rcl}
C^{\pm\pm}_{ijrs}&=&\sum\limits_{x\in\Om}C^{\pm\pm}_{ijrs}(x)\\
\\
H&=&\sum\limits_{x\in\Om}H(x).
\era
\eeq

In the result (26), we remark that the hermetian operator $H(x)$ looks like
a deformed supersymmetry Hamiltonian operator describing a peculiar
particles constructed from $k_i$-fermions and defined on 2d space. The
deformation, in this case, looks be normal since the basis of our construction
is deformed system ($k_i$-fermionic one) which generalizes the bosonic and
fermionic ones and also the presence of special topological effects of 2d
space in which anyons live.
\section{Irreducible Representations of anyonic algebras and supralgebra}
\hspace{.3in}To construct the representations of anyonic algebras treated
above, we will consider a Fock space. In first of all, let us give a local
irreducible representation on a Fock space of $k_i$-fermionic algebra. We
introduce this space of this algebra by the set
\beq
F_{i_{x}}=\lbrace\vert n_{i_{x}}\rangle, n_{i_{x}}=0,1,...,k_{i}-1\rbrace
\eeq
where the notation $i_{x}$ means that this Fock space is introduced in
each site $x$ of the lattice $\Omega$.

The action of $k_i$-fermionic operators $f_{i}^- (x)$ and $f_{i}^{+}(x)$ on $F_{i_{x}}$ is
expressed by the following equalities
\beq
\bra{cc}
f_{i}^{+}(x)\vert n_{i_{x}}\rangle=\vert n_{i_{x}}+1\rangle,& f_{i}^{+}(x)\vert k_{i}-1\rangle=0, \\
\\
f_{i}^- (x)\vert n_{i_{x}}\rangle=[n_{i_{x}}]_{q_{i}}\vert n_{i_{x}}-1\rangle,& f_{i}^- (x)\vert 0\rangle=0.
\era
\eeq
Then, the operator $f_{i}^{+}(x)$ is called a creation $k_i$-fermionic operator on the site $x$ and $f_{i}^- (x)$ is an annihilation one. These generators also
satisfy the following nilpotency condition which is coherent with the above
equalities. So, we have
\beq
(f_{i}^- (x))^{k_{i}}=(f_{i}^{+}(x))^{k_{i}}=0
\eeq
which generalizes the Pauli exclusion principle; i.e. we can not
find in one state more than $k_{i}-1$ particles of the $i^{th}$ kind.

Owing to the definition of anyonic operator given by Eq.(8) the irreducible
representation space is the same one of $k_i$-fermionic system; $F_{i_{x}}$. Thus, we can prove that the algebraic relations of Eq.(11) are coherent with the
action of anyonic operators $a^{\pm}_{ij}(x_{\pm})$ on the Fock space
$F_{i_{x}}$. We get
\beq
\bra{rcl}
a^{+}_{ij}(x_{\pm})\vert n_{i_{x}}\rangle&=&e^{i\frac{\nu_{ij}}{2}\sum\limits_{y\neq x} \Te_{\pm\Ga_{x}}(x,y)}\vert n_{i_{x}}+1\rangle \\
\\
a_{ij}^- (x_{\pm})\vert n_{i_{x}}\rangle&=&[n_{i_{x}}]_{q_{i}}e^{-i\frac{\nu_{ij}}{2}\sum\limits_{y\neq x} \Te_{\pm\Ga_{x}}(x,y)}\vert n_{i_{x}}-1\rangle \\
\\
a^{+}_{ij}(x_{\pm})\vert k_{i}-1\rangle&=&0 \\
\\
a_{ij}^- (x_{\pm})\vert 0\rangle&=&0.
\era
\eeq
According to these relations, we see the operators $a_{ij}^+ (x_{\pm})$ and
$a^{-}_{ij}(x_{\pm})$ as a creation and annihilation anyonic operators
respectively.

Let us now define the Fock-like space of combined anyonic system as a direct sum of Fock spaces $F_{i_{x}}$ defined in the equality (29). We write
\beq
F_{x}=\bigoplus\limits_{i=0}^{n-1}F_{i_{x}}.
\eeq
In this space, the vacuum state and $n$-particles state are
described on each $x$ on the lattice $\Om$, and denoted, respectively, by
\beq
\bra{rl}
(\vert 0\rangle)_{i}\equiv \pmatrix{\vert 0\rangle_{i}\cr
\vert 0\rangle_{i+1}\cr
\vdots\cr
\vert 0\rangle_{i+n-1}\cr},& (\vert n\rangle)_{i}\equiv \pmatrix{\vert n\rangle_{i}\cr
\vert n\rangle_{i+1}\cr
\vdots\cr
\vert n\rangle_{i+n-1}\cr}
\era
\eeq
with index $i=0.1,...,n-1$ of components in cyclic order. We remark, owing to equation (32), that the action of $A_{ij}(x_{\al})$ and $A_{ij}^{\dag}(x_{\al})$, defined in equations (13) on $F_x$ will be given by
\beq
\bra{rl}
A_{ij}^- (x_{\pm}) (\vert 0\rangle)_{i}=0, & A_{ij}^{+}(x_{\pm}) (\vert
k_{i}-1\rangle)_{i}=0\\
\\
A_{ij}^- (x_{\pm}) (\vert n\rangle)_{i}= A (\vert n-1\rangle)_{i}, &
A_{ij}^{+}(x_{\pm}) (\vert n\rangle)_{i}= B(\vert n+1\rangle)_{i}
\era
\eeq
with
\beq
\bra{ll}
A = \pmatrix{[n]_{q_{i}}e^{-i\frac{\nu_{ij}}{2}\sum\limits_{y\ne x}\Te_{\pm\Ga_{x}}(x,y)}&&&\cr
& [n]_{q_{i+1}}e^{-i\frac{\nu_{i+1,j}}{2}\sum\limits_{y\ne x}\Te_{\pm\Ga_{x}}(x,y)}&&\cr
&&\ddots&\cr
&&& [n]_{q_{i+n-1}}e^{-i\frac{\nu_{i+n-1,j}}{2}\sum\limits_{y\ne x}\Te_{\pm\Ga_{x}}(x,y)}\cr} \\
\\
B = \pmatrix{e^{i\frac{\nu_{ij}}{2}\sum\limits_{y\ne x}\Te_{\pm\Ga_{x}}(x,y)} &&&\cr
& e^{i\frac{\nu_{i+1,j}}{2}\sum\limits_{y\ne x}\Te_{\pm\Ga_{x}}(x,y)}&&\cr
&&\ddots&\cr
&&&e^{i\frac{\nu_{i+n-1,j}}{2}\sum\limits_{y\ne x} \Te_{\pm\Ga_{x}}(x,y)}\cr}.
\era
\eeq
\vskip.2cm
Then, the relations of Eq.(35) are compatible with the nilpotency condition
Eq.(15), and we can call $A_{ij}^- (x_{\pm})$ annihilation operator and
$A_{ij}^+ (x_{\pm})$ creation one.

Now, let us give the representation of the supersymmetry constructed on 2d
lattice. The action of the supercharges $C^{\pm -}_{ijrs}(x)$ on the
associated Fock-like space that we define as
\beq
F=\bigoplus\limits_{\small \bra{rcl}
i,r=0\\ i\ne r
\era
}^{n-1}(F_{i_{x}}\otimes F_{r_{x}})
\eeq
is given by the equalities
\beq
\bra{rcl}
C^{\pm -}_{ijrs}(x) (\vert 0\rangle)_{i}\ot (\vert 0\rangle)_{r}&=&0\\ \\
C^{+-}_{ijrs}(x) (\vert n\rangle)_{i}\ot (\vert n\rangle)_{r}&=&C (\vert
n-1\rangle)_{i}\ot (\vert n+1\rangle)_{r} \\
\\
C^{--}_{ijrs}(x) (\vert n\rangle)_{i}\ot (\vert n\rangle)_{r}&=&D (\vert
n+1\rangle)_{i}\ot  (\vert n-1\rangle)_{r}\\ \\
C^{\pm -}_{ijrs}(x) (\vert k_i -1\rangle)_{i}\ot (\vert k_i -1\rangle)_{r}&=&0
\era
\eeq
and their hermitian conjugate $C^{\pm +}_{ijrs}(x)$ act on the Fock like-space
\beq
F'=\bigoplus\limits_{\small \bra{rcl}
r=0\\ i\ne r
\era}^{n-1}(F_{r_{x}}\otimes F_{i_{x}})
\eeq
as follows
\beq
\bra{rcl}
C^{\pm +}_{ijrs}(x) (\vert 0\rangle)_{r}\ot (\vert 0\rangle)_{i}&=&0\\ \\
C^{-+}_{ijrs}(x) (\vert n\rangle)_{r}\ot (\vert n\rangle)_{i}&=&E (\vert n-1\rangle)_{r}\ot (\vert n+1\rangle)_{i}\\
\\
C^{++}_{ijrs}(x)(\vert n\rangle)_{r}\ot (\vert n\rangle)_{i}&=&F (\vert n+1\rangle)_{r}\ot (\vert n-1\rangle)_{i}\\ \\
C^{\pm +}_{ijrs}(x) (\vert k_i -1\rangle)_{r}\ot (\vert k_i -1\rangle)_{i}&=&0

\era
\eeq
where
$$C=\pmatrix{[n]_{q_i}e^{\frac{i}{2}\sum\limits_{y\ne x}(\nu_{rs}\Te_{+\Ga_x}(x,y)-\nu_{ij}\Te_{-\Ga_x}(x,y))} &&\cr
&\ddots&\cr
&&[n]_{q_{i+n-1}}e^{\frac{i}{2}\sum\limits_{y\ne x}(\nu_{r+n-1,s}\Te_{+\Ga_x}(x,y)-\nu_{i+n-1,j}\Te_{-\Ga_x}(x,y))}\cr}$$
\\
$$D=\pmatrix{[n]_{q_r}e^{\frac{i}{2}\sum\limits_{y\ne x}(\nu_{ij}\Te_{+\Ga_x}(x,y)-\nu_{rs}\Te_{-\Ga_x}(x,y))} &&\cr
&\ddots&\cr
&& [n]_{q_{r+n-1}}e^{ \frac{i}{2}\sum\limits_{y\ne x}(\nu_{i+n-1,j}\Te_{+\Ga_x}(x,y)-\nu_{r+n-1,s}\Te_{-\Ga_x}(x,y))}\cr}$$
\\
$$E=\pmatrix{[n]_{q_r}e^{\frac{i}{2}\sum\limits_{y\ne x}(\nu_{ij}\Te_{-\Ga_x}(x,y)-\nu_{rs}\Te_{+\Ga_x}(x,y))} &&\cr
&\ddots&\cr
&& [n]_{q_{r+n-1}}e^{\frac{i}{2}\sum\limits_{y\ne x}(\nu_{i+n-1,j}\Te_{-\Ga_x}(x,y)-\nu_{r+n-1,s}\Te_{+\Ga_x}(x,y))}\cr}$$
\\
$$F=\pmatrix{[n]_{q_i}e^{-\frac{i}{2}\sum\limits_{y\ne x}(\nu_{ij}\Te_{+\Ga_x}(x,y)-\nu_{rs}\Te_{-\Ga _x}(x,y))} &&\cr
&\ddots&\cr
&& [n]_{q_{i+n-1}}e^{-\frac{i}{2}\sum\limits_{y\ne x}(\nu_{i+n-1,j}\Te_{+\Ga_x}(x,y)-\nu_{r+n-1,s}\Te_{-\Ga_x}(x,y))}\cr}$$

Owing to these results the irreducible representations of anyonic algebras
were considered and it was shown that they are related to generalized Pauli
exclusion principle. Furthermore, we could see the supersymmetry of our
combined system has deformation properties plus "fractional" properties
coming from the basis particles ($k_i$-fermions) and the nature of anyons
respectively.

\section*{Concluding remarks}
\hspace{.3in}To summarize, exotic statistics were introduced in physics as
an exotic extension of bosonic and fermionic statistics, and the both statistics
could be unifiedd by SUSY. Recently, the existence of intimate relation
between exotic statistics and SUSY was established by observation of
hidden SUSY structure in purely  parabosonic and purely parafermionic
systems. So, the SUSY and exotic statistics can be unified in the form of
paraSUSY for parafermions, and also in the form of so-called fractional SUSY for
$q$-bosons where $q$ is a root of unity and nonlinear SUSY for bosonization of supersymmetric quantum mechanics. These studies were formulated on four dimensional spacetime.

For lower dimensions, we have discussed in this work what we could call
fractional SUSY in a general case for $k_i$-fermionic anyons. As generalization, our work was for unification of different exotic statistics on 2d space. These kinds of statistics describe quasi-particles those we defined as deformed particles constructed from bosons or fermions. To generalize this construction we have considered $k_i$-fermions as basis to built our generalized anyons, those we call $k_i$-fermionic anyons. From our present construction, it is easy to remark that in the limit cases $k_i=2$ we refind the anyonic oscillators defined in the reference \cite{20} by Lerda and Sciuto, and for $k_i \longrightarrow\infty$ our anyons go to bosonic anyons.

These results are compatible with the fact that anyonic particles having
different fractional charge and spin can not live peacefully, such that the
commutation relations for anyonic operators show the generalization of
exclusion principle and the interchange of two anyons can not be defined
consistently for them. These analyses are very interesting for various fields as
condensed matter in which the experimental advances related to the fractional
quantum Hall effect have proved that quasi-particles discovered appear to
exhibit anyonic behavior. Also, the field of quantum compuattion which
can be constructed from the abstract study of anyonic systems, such that the
braiding and fusion of anyonic excitations in quantum Hall electron liquids
and 2D magnets are modeled by modular functors opening a new possibility
for the realization of quantum computers \cite{25,26}.

\section*{Acknowledgments}
\hspace{.3in}The author would like to thank the Abdus Salam ICTP, particularly the associateship Sckeme for its warm hospitality, and the Max-Planck-Institut f\"ur Physik Komplexer Systeme for link hospitality during the stage in which one part of this paper was done.

\end{document}